\documentclass[12pt]{iopart}
\usepackage[utf8]{inputenc}
\usepackage{iopams}
\usepackage{graphicx}
\usepackage{url}
\usepackage{cleveref} 
\usepackage{color}
\usepackage[dvipsnames]{xcolor}

\newcommand{\enm}{E^{\rm NM}}
\newcommand{\rhosat}{\rho_{\rm sat}}
\newcommand{\knm}{K^{\rm NM}}
\newcommand{\asym}{a_{\rm sym}^{\rm NM}}
\newcommand{\lsym}{L_{\rm sym}^{\rm NM}}

\newcommand{\msca}{m_s^{*}}
\newcommand{\mvec}{m_v^{*}}
\newcommand{\CrDr}[1]{C_{#1}^{\rho\Delta\rho}}
\newcommand{\VZeroN}{V_0^n}
\newcommand{\VZeroP}{V_0^p}
\newcommand{\VZeroq}{V_0^q}
\newcommand{\CrDJ}[1]{C_{#1}^{\rho\nabla J}}

\newcommand{\POUNDERS}{\textsc{pounders}}

\newcommand{\HFBTHO}{\textsc{hfbtho}}

\newcommand{\UNEDF}{\textsc{unedf}}
\newcommand{\UNEDFZERO}{\textsc{unedf0}}
\newcommand{\UNEDFONE}{\textsc{unedf1}}
\newcommand{\UNEDFTWO}{\textsc{unedf2}}

\newcommand{\tensor}[1]{\mathsf{#1}}
\newcommand{\gras}[1]{\boldsymbol{#1}}

\newcommand{\beqn}{\begin{eqnarray}}
\newcommand{\eeqn}{\end{eqnarray}}
\newcommand{\beqnu}{\begin{eqnarray}}
\newcommand{\eeqnu}{\end{eqnarray}}
\newcommand{\be}{\begin{equation}}
\newcommand{\ee}{\end{equation}}
\newcommand{\beu}{\begin{equation*}} 
\newcommand{\eeu}{\end{equation*}} 
\newcommand{\ba}{\begin{array}}
\newcommand{\ea}{\end{array}}

\newcommand{\rms}{r.m.s.}

\newcommand{\DFT}{\textsc{dft}}
\newcommand{\HF}{\textsc{hf}}
\newcommand{\BCS}{\textsc{bcs}}
\newcommand{\HFB}{\textsc{hfb}}
\newcommand{\EDF}{\textsc{edf}}

\newcommand{\fb}{\mathbf{f}}

\newcommand{\nub}{\gras{\nu}}

\newcommand{\xb}{\mathbf{x}}

\newcommand{\hyperp}{\gras{\kappa}}

\newcommand{\nd}{n_d} 
\newcommand{\nx}{n_x} 
\newcommand{\nnuc}{n_{\rm nuc}} 
\newcommand{\ndes}{D} 

\newcommand{\mcmc}{\textsc{mcmc}}
\newcommand{\gp}{\textsc{gp}}

\newcommand{\qtarget}{Q_{\rm t}}
\newcommand{\qfinal}{Q_{\rm f}}
\newcommand{\pes}{\textsc{pes}}

\begin{document}

\begin{flushright}\mbox{LLNL-JR-805357}\end{flushright}

\title[EDF Calibration with Deformed Nuclei]{Calibration of Energy Density Functionals with Deformed Nuclei}

\author{N. Schunck,$^1$ J. O'Neal,$^2$ M. Grosskopf,$^3$ E. Lawrence,$^3$ S.M.~Wild$^2$}

\address{$^1$ Nuclear and Chemical Science Division, Lawrence Livermore National Laboratory, Livermore, CA 94551, USA}
\address{$^2$ Mathematics and Computer Science Division, Argonne National Laboratory, Lemont, IL 60439, USA}
\address{$^3$ Computer, Computational, and Statistical Sciences Division, Los Alamos National Laboratory, Los Alamos, NM 87545, USA}
\ead{schunck1@llnl.gov}

\begin{abstract}
Nuclear density functional theory is the prevalent theoretical framework for
accurately describing nuclear properties at the scale of the entire chart of
nuclides. Given an energy functional and a many-body scheme (e.g., single- or
multireference level), the predictive power of the theory depends strongly on
how the parameters of the energy functionals have been calibrated with
experimental data. Expanded algorithms and computing power have enabled recent
optimization protocols to include data in deformed nuclei in order to optimize
the coupling constants of the energy functional. The primary motivation of this
work is to test the robustness of such protocols with respect to some of the
technical and numerical details of the underlying calculations, especially when
the calibration explores a large parameter space. To this end, we quantify the
effect of these uncertainties on both the optimization and statistical
emulation of composite objective functions. We also emphasize that Bayesian
calibration can provide better estimates of the theoretical errors used to
define objective functions.
\end{abstract}

\vspace{2pc}

\submitto{\JPG}

\section{Introduction}
\label{sec:intro}

Nuclear theory plays an essential role in many fundamental science problems
\cite{nsaccommittee2015}. In particular, it provides data for simulations of
the origin of the elements in astrophysical environments, particularly the
rapid-neutron capture process, which involves very neutron-rich, short-lived
nuclei for which  no experimental measurements exist \cite{mumpower2016}.
Current research on a possible end for the periodic table of elements also
involves advanced nuclear calculations for superheavy elements, where fission
plays a major role \cite{giuliani2019}. Tests of fundamental symmetries and the
search for physics beyond the Standard Model also often depend on
high-accuracy, high-precision calculations of nuclear properties
\cite{engel2003,dobaczewski2005a}.

A common feature of all theoretical nuclear models is that they are imperfect.
In this work, we focus on nuclear energy density functional theory ({\DFT}),
which relies on an effective description of nuclear forces encapsulated in the
form of an energy density functional ({\EDF}) and is the prevalent framework
for computing heavy nuclei \cite{schunck2019}. Because of the disconnect
between realistic nuclear potentials and the effective encoding of many-body
effects in the functional, {\DFT} should be viewed as a phenomenological model
with unknown parameters that must be calibrated with a set of experimental data
(see, e.g., \cite{niksic2008a,klupfel2009,kortelainen2010,kortelainen2012,
erler2013,chen2014,kortelainen2014,navarroperez2018}). This naturally induces
uncertainties and errors that have been extensively discussed in the literature \cite{dobaczewski2014,schunck2015,schunck2015a,schunck2015c}.

Traditionally, energy functionals were often fitted to nuclear matter
properties together with a small sample of properties in doubly magic,
closed-shell nuclei (see, e.g., \cite{beiner1975,chabanat1997,chabanat1998,
brown1998}). In addition to the advantage in computational cost, this strategy
was often justified by the fact that (i) in such nuclei, pairing correlations
(e.g., as described by Bardeen–Cooper–Schrieffer ({\BCS}) or the
Hartree-Fock-Bogoliubov ({\HFB}) theory) automatically collapse: the parameters
of the pairing functional (particle-particle channel) are thus decoupled from
the ones of the mean field (particle-hole channel), and (ii) the shell effects
that determine many deformation properties of nuclei originate from the
spontaneous symmetry breaking of rotational invariance: if the spherical shell
structure is properly reproduced, deformed shell gaps will automatically appear
for the correct number of particles. The success of ``single-particle
phenomenology'' in describing broad swaths of nuclear properties gave credence
to this approach; see, for example, \cite{nilsson1995} for an overview.

As traditional computational bottlenecks in {\DFT} applications slowly
disappear, this na\"ive approach to {\EDF} calibration has been questioned
\cite{kortelainen2010}. In particular, the importance of the spherical shell
structure can be nuanced for at least three reasons. First, single-particle
levels are not experimental observables \cite{duguet2015}: not only are they
extracted from experimental data in a model-dependent way, but for all except
the Hartree-Fock ({\HF}) theory they cannot be unambiguously related to actual
observables of the model \cite{schunck2019,duguet2012}. Second, several studies
have shown that correlations beyond {\HF} (e.g., particle-vibration couplings)
have a large impact on such shell structure \cite{colo2010,cao2014a,
tarpanov2014a}: forcing a fit at the {\HF} level will thus cause overfitting.
Misfits are unavoidable, and it is thus highly unlikely that one could
reproduce exactly a given shell structure (a problem recognized also in nuclear
phenomenology; see \cite{dudek2011,szpak2011,dudek2013}). Third, deformation
properties are the result of a competition between shell and bulk (liquid-drop)
effects: the analyses of \cite{bender2006,nikolov2011} show that bulk surface
properties, particularly isovector ones, can  be constrained only by
calculations in very deformed nuclei. Performing a fit in deformed nuclei thus
seems unavoidable, since it reflects the fact the {\HFB} theory is an imperfect
model that must be carefully calibrated with all types of observables that fall
within its scope.

In the important case where fitting involves deformed {\HF} or {\HFB}
calculations,  careful examination is warranted to determine how the
self-consistent iterative process is initialized. In other applications, such
as fission, these initial conditions are known to be crucial \cite{dubray2012}:
for a given parameterization of the functional, does every {\HFB} calculation
in the fit converge to an appropriate value? In other words, are all deformed
nuclei really deformed? Are fission isomers truly separated from the ground
state by a barrier? Such questions can be especially relevant for methods based
on supervised learning, where the training of the model involves exploring a
large section of the parameter space, some regions of which may lead to
nonphysical results.

The goal of this paper is thus to study the robustness of the optimization and
calibration protocols that include deformed nuclei. Specifically, we focus on
the calibration of the Skyrme {\UNEDF1} functional. We seek to (i) quantify the
impact of changes in initial conditions for the underlying {\HFB} calculations
on this function (i.e, the ``forward model''); (ii) analyze the behavior of an
optimization algorithm for the inverse problem under such changes; and (iii)
quantify the impact on the training of statistical models in the context of
Bayesian calibration.

In \Cref{sec:background} we review some basic elements of nuclear density
functional theory, the {\UNEDFONE} functional, and the optimization software
based on the {\HFBTHO} solver. In \Cref{sec:casestudy} we detail the case study
considered here, which varies the deformation of the initial state in {\HFB}
calculations while keeping conditions such as the dataset, platform, and
parameter set constant. In \Cref{sec:casestudy} we also study the effect of
these changes on the forward computation and optimization-based solution of the
inverse problem. In \Cref{sec:calibration} we study both the effect on
statistical emulation with Gaussian processes ({\gp}s) and the downstream effect
on {\gp}-based calibration.


\section{Theoretical and Computational Background}
\label{sec:background}

The general physics framework for this and all previous {\UNEDF} work is the
{\HFB} theory, where the nuclear many-body wave function has the form of a
quasiparticle vacuum; see \cite{schunck2019} for a review.


\subsection{The Hartree-Fock-Bogoliubov Theory with Skyrme Generators}
\label{sec:hfb}

In the {\HFB} theory, the one-body density matrix and pairing tensor are the
main degrees of freedom. The total energy of the nucleus at the {\HFB}
approximation can thus be expanded as
\be
E[\rho,\kappa,\kappa^{*}] = E_{\rm nuc}[\rho] + E_{\rm Cou}[\rho] + E_{\rm pair}[\kappa,\kappa^{*}].
\label{eq:etot}
\ee
For the nuclear part of the energy functional (\ref{eq:etot}), we consider a
Skyrme-like {\EDF},
\beu
E_{\rm nuc}[\rho] = \sum_{t=0,1} \int d^{3}\gras{r}\; \chi_t(\gras{r}),
\eeu
where the functional includes the kinetic energy term and is expressed as
\beu
\chi_t(\gras{r}) =
  C_t^{\rho\rho} \rho_t^2
+ C_t^{\rho\tau} \rho_t\tau_t
+ C_t^{JJ} \tensor{J}^{2}_t
+ C_t^{\rho\Delta\rho} \rho_t\Delta\rho_t
+ C_t^{\rho \nabla J} \rho_t\gras{\nabla}\cdot\gras{J}_t.
\eeu
Here, the index $t$ refers to the isoscalar ($t=0$) or isovector $(t=1$)
channel. The definitions of the various densities $\rho$, $\tau$, and
$\tensor{J}$ ($\gras{J}$ is the vector part of $\tensor{J}$) can be found in
\cite{engel1975,dobaczewski1996,bender2003,perlinska2004,lesinski2007}. The
parameters of the model are the coupling constants $C_{t}^{uu'}$, all of which
are real-valued scalars with the exception of $C_t^{\rho\rho}$, which has a
density dependency of the form
\beu
C_t^{\rho\rho} =  C_{t0}^{\rho\rho}  + C_{t{\rm D}}^{\rho\rho}~ \rho^{\gamma}_{0}(\gras{r}).
\eeu
The full description of the particle-hole channel requires 13 parameters.

The Coulomb term in (\ref{eq:etot}) is computed at the {\HF} approximation with
the exchange term treated with the Slater approximation \cite{bender2003}. The
pairing energy is computed at the {\HFB} approximation with an approximate
Lipkin-Nogami correction based on a simple seniority pairing force; see
\cite{stoitsov2003,stoitsov2007} for details. The pairing  functional itself
originates from a surface-volume density-dependent pairing force
\beu
V_{q}(\gras{r},\gras{r}') =
\VZeroq\left[ 1 - \frac{1}{2}\frac{\rho(\gras{r})}{\rho_{c}} \right ]
\delta(\gras{r}-\gras{r}'),
\eeu
where $q$ indicates the type of particle (proton or neutron) and
$\rho_{c} = 0.16 $ fm$^{-3}$. Including the pairing channel in the fit thus
adds two more parameters, resulting in a total of 15 parameters.


\subsection{The UNEDF1 Energy Functional}
\label{sec:unedf1}

In the {\UNEDF} optimization protocol described in \cite{kortelainen2010,
kortelainen2012,kortelainen2014}, the coupling constants $C_{t0}^{\rho\rho}$,
$C_{t{\rm D}}^{\rho\rho}$, $\gamma$, and $C_t^{\rho\tau}$ are expressed as a
function of the parameters of infinite nuclear matter \cite{kortelainen2010}.
As a result, the vector $\xb$ of parameters that can be adjusted in {\UNEDF}
fits is
\beqnu
\fl
\left(
\enm, \rhosat, \knm, \asym, \lsym, 1/\msca, 1/\mvec,
\CrDr{0}, \CrDr{1}, \CrDJ{0}, \CrDJ{1}, C_0^{JJ}, C_1^{JJ},
\VZeroN, \VZeroP
\right).\nonumber
\eeqnu
In all {\UNEDF} fits, the vector effective mass was kept constant at the SLy4
value of $\mvec = 1.249 838$; see \cite{chabanat1998} for details. In the
{\UNEDFZERO} and {\UNEDFONE} fits, the tensor coupling constants were set to 0
(i.e., $C_{0}^{JJ} = C_{1}^{JJ} = 0$), reducing the number of fit parameters to
12.

The $\chi^2$ criterion that defines an optimization's objective function and
enters the expression for the likelihood is
\be
\chi^2(\xb) = \frac{1}{n_d - n_x} \sum_{i=1}^{D_{T}} \sum_{j=1}^{n_i}
\left(
\frac{s_{ij}(\xb) - d_{ij}}{\sigma_i}
\right)^{2},
\label{eq:chi2}
\ee
where $n_x$ is the number of fit parameters, $D_T$ is the number of data types,
$n_i$ is the number of data points for the data type $i$, $n_d$ is the total
number of data points (i.e., $n_d=\sum_{i=1}^{D_{T}} n_i$),  $s_{ij}(\xb)$ is
the simulation output for the point $j$ of data type $i$, $d_{ij}$ is the
corresponding experimental value, and $\sigma_i$ the estimate of the error for
data type $i$.  We recall that {\UNEDFONE}, the paradigm studied below, had the
following characteristics.
\begin{itemize}
\item $n_x = 12$ for the original {\UNEDFONE} fit in \cite{kortelainen2012};
\item $D_T = 4$ with binding energies $(i=1)$, proton {\rms} radii ($i=2$),
odd-even staggering (OES) energy ($i=3$), and excitation energy of fission
isomers ($i=4$);
\item $n_d = 115$---see supplemental material of \cite{kortelainen2014} for
details of the nuclei included;
\item $\sigma_1 = 2$ MeV, $\sigma_2 = 0.02$ fm, $\sigma_3 = 0.05$ MeV,
and $\sigma_4 = 0.5$ MeV, for masses, proton radii, OES, and fission isomers,
respectively.
\end{itemize}
In contrast to the original {\UNEDFONE} paper \cite{kortelainen2012}, we  also
use the AME 2016 mass evaluation \cite{wang2017} for all binding energies. To
extract nuclear binding energies from results tabulated in the mass evaluation,
we subtract the electronic binding energy for which we use the empirical
formula
\beu
B_e(Z) = 1.44381\times 10^{-5}Z^{2.39} + 1.55468\times 10^{-12} Z^{5.35},
\eeu
with the energy given in MeV. We include only true experimental measurements
and do not take into account evaluated masses.


\subsection{The HFBTHO Solver}
\label{sec:hfbtho}

Our {\DFT} solver was based on the latest version of the {\HFBTHO} program
\cite{perez2017}.  {\HFBTHO} solves the {\HFB} equation by expanding the
solutions in the harmonic oscillator basis and by using successive
diagonalizations of the {\HFB} matrix until convergence (within a numerical
tolerance) is achieved. Throughout this manuscript, we will refer to a
\emph{nuclear configuration} as a {\HFB} solution for a nucleus with $Z$
protons and $N$ neutrons corresponding to a local minimum of the potential
energy curve as a function of the axial quadrupole moment.
For a single nuclear configuration and a given point
$\xb$ in the functional's $\nx$-dimensional parameter space, {\HFBTHO}
calculates theoretical observable values $s_{ij}(\xb)$ for the binding energy,
proton {\rms} radius, proton pairing gap, and neutron pairing gap entering the
objective function (\ref{eq:chi2}). For a given parameterization $\xb$, the
resulting value of the objective function is contingent on both the precision
and the accuracy of the underlying {\HFB} calculation.
\begin{itemize}
\item The \emph{numerical precision} depends on a number of ``hyperparameters''
characterizing the basis (e.g., oscillator length $b_0$, basis deformation
$\beta_{\rm HO}$, number of shells $N_0$, number of states $N_{\rm states}$),
and  quadrature grid (number of points for Hermite, Laguerre, and Legendre
quadratures). In this work, we adopt the same conventions as in
\cite{kortelainen2010,kortelainen2012,kortelainen2014} concerning the basis
characteristics. Specifically, we set $b_{0} = \sqrt{\hbar/m\omega_{0}}$ using
$\omega_{0} = 1.2\times 41/A^{1/3}$ (see, e.g., \cite{stoitsov2013}),
$N_0 = 20$, $\beta_{\rm HO} = 0$ for all ground-state calculations, and
$\beta_{\rm HO} = 0.4$ for the fission isomer calculations. We also set
$N_{\rm Her} = 40$, $N_{\rm Lag} = 40$, and $N_{\rm Leg} = 80$ for the Hermite,
Laguerre, and Legendre quadratures, respectively.
\item The \emph{physics accuracy} depends on the characteristics of the initial
condition used to start the {\HFB} iterations. In {\HFBTHO}, iterations are
initialized with the solution of the Schr\"odinger equation for a Woods-Saxon
potential with quadrupole, octupole, and hexadecapole deformation. In practice,
one often specifies only the quadrupole deformation $\beta_{2}$ of the
Woods-Saxon potential based on the expected value of the mass quadrupole moment
of the nucleus, which we denote $\qtarget$. For a given configuration,
parameter point $\xb$, and nucleus ($Z, N$), the final quadrupole moment (i.e.,
at approximate convergence) is denoted by $\qfinal \equiv \qfinal(Z,N,\xb)$.
\end{itemize}

In performing optimization or Bayesian calibration, one may explore a large
area of the $\nx$-dimensional parameter space; there is no guarantee that the
{\HFBTHO} program converges for all queried $\xb$. In {\HFB} iteration $n$,
the convergence metric defined as the Euclidean norm $\mu^{(n)} = \|\rho^{(n)}-\rho^{(n-1)}\|_{2} =
\sqrt{\sum_{ij} \big[\rho_{ij}^{(n)}-\rho_{ij}^{(n-1)}\big]^2}$ is a  simple
filter for accepting the result of the calculation: if $\mu^{(n)} > \varepsilon$ 
after $n=500$ iterations, the results are immediately discarded. In this work, we set $\varepsilon = 10^{-5}$.

Even when the calculation passes this filter (i.e., $\mu \leq \varepsilon$),
the result may not be physically correct. A first layer of offline
postprocessing is thus responsible for flagging as nonsensical those results
that have at least one theoretical observable value that is ``too far'' from
physical expectations. Specifically, we flag solutions that have a pairing gap
less than -10 keV; a binding energy per nucleon of $E/A < -11$ MeV (either in
the ground state or in the fission isomer); or a proton radius $r_{p}$ outside
of the interval $[0.8, \,  1.1] A^{1/3}$.

A second stage of flagging is applied to those results that have potentially
sensible observable values but for which the values are not consistent with the
expected characteristics of the configuration of interest. This stage  consists
mostly of identifying abnormal values of the final quadrupole deformation
$\beta_{2}$. For example, a spherical ground state is flagged if its axial
quadrupole deformation has $|\beta_{2}| > 0.01$; a deformed ground state is
flagged unless $0.05 \le \beta_{2} \le 0.6$; and a fission isomer configuration
is flagged unless $0.3 \le \beta_{2} \le 1.15$. In addition, we  require that
the fission isomer final deformations  be sufficiently larger than the
associated ground-state deformation so that  a potential barrier can exist
between the two states. This requirement is enforced by insisting that all
valid results  satisfy $\beta_{2}^{(\rm FI)} > 2.7 \beta_{2}^{(\rm g.s.)}$.

Fission isomers are excited states; therefore their binding energy $E_{\rm FI}$
should exceed that of the associated deformed ground-state $E_{\rm g.s.}$.
While we may allow the fission isomer to be lower than the ground state for
some very neutron-rich or superheavy nuclei, our dataset does not contain such
exotic systems.
Therefore,
in addition to tagging fission isomer and deformed ground-state data as
nonsensical or nonphysical because of their final deformation, we  apply a
final layer of outlier analysis. Specifically, a valid fission isomer result
that has a valid associated deformed ground-state result is tagged as
nonphysical if $E_{\rm FI} < E_{\rm g.s.} - 0.5$ MeV.


\subsection{The Observable Engine}
\label{sec:obs_engine}

Each of our studies is defined with respect to a set of $\nd$ observables that
are associated with $\nnuc$ distinct nuclear configurations; for {\UNEDFONE},
$\nnuc = 79$. Our studies require information about how each observable's
theoretical values vary across the parameter space.  For instance, in order to
find approximate minimizers of an objective function, optimization software
assembles theoretical results of these observables, one parameter space point
at a time,  to determine the next parameter point for evaluation.

To help acquire and manage such potentially large amounts of data, we developed
on top of the {\HFBTHO} solver \cite{perez2017} a layer of software
parallelized with MPI, which we call the \emph{observable engine}. This
software uses the output of {\HFBTHO} to generate and gather theoretical
observable results at each configuration and parameter space point combination
contained in the Cartesian product of the $\nnuc$ nucleus configurations for a
given set of parameter space points.  The observable engine also performs
online postprocessing of the {\HFBTHO} results to compute all derived
theoretical observable values (e.g., fission isomer excitation energies). In the
case of optimization, the observable engine is called for each new parameter
space point $\xb$. In the case of the Bayesian study reported in
Section~\ref{sec:calibration}, the observable engine was run on a design
$\mathcal{D}$ of $\ndes$ distinct parameter space points,
$\mathcal{D} = \{ \xb_{i} \}_{i=1,\dots,\ndes}$; in the study in
Section~\ref{sec:calibration}, we used $\ndes=500$.


\section{Case Study and Impact of Target Quadrupole Moment on Forward Model Calculations and Optimization-Based Inversion}
\label{sec:casestudy}

Self-consistent {\HFB} calculations can be unstable with respect to how they
are initialized. This problem is relevant for all deformed {\HFB} calculations
and is well known by fission practitioners in the calculation of potential
energy surfaces ({\pes}s) since it leads to infamous discontinuities
\cite{dubray2012}. For calibration purposes, these instabilities must be kept
tightly under control: in the case of the excitation energy $E^{*}$ of the
fission isomer, for example, one cannot accept that two slightly different
values of the quadrupole moment for the initial density lead to significantly
different results for $E^{*}$. In this work, we focus on the impact of the
value $\qtarget$ of the quadrupole moment used to drive the self-consistent
iterations toward a given solution, either through the determination of the
initial deformation or as a constraint imposed on the first few iterations.

Ideally, any ambiguity could be avoided if, for a given nuclear configuration
and parameter space point $\xb$, we would estimate the local {\pes} over the
range of physically realistic final quadrupole moments, identify all local
extrema, and use some physics-based criterion to identify which local minimum
is the correct solution. For a study with a large number of configurations and
parameter space points, however, this brute-force method rapidly becomes
infeasible. In addition, when Lipkin-Nogami corrections are activated (as is
the case with the {\UNEDF1} functional), the self-consistent solution is no
longer variational: selected local minima over constrained values of quadrupole
moments may not correspond to unconstrained minima.

In the following subsections, we describe a data acquisition, configuration,
and analysis scheme that allows one to acquire data without computing local
{\pes}s and to determine whether choosing only one out the many possible
solutions that {\HFBTHO} can find unduly affects the outcome of the analysis.
To this end, we collect the same dataset with multiple $\qtarget$
configurations. For a given nuclear configuration computed at a given parameter
space point $\xb$, the computation is assumed to be \emph{independent} of
$\qtarget$ if each of the computations run with a different value of $\qtarget$
converges, if all observables are physically reasonable (i.e., not flagged by
the rules stated in \Cref{sec:hfbtho}), and if the results are effectively
identical across all target configurations. Because of the lack of variability,
such a result would suggest (without proving) that there is a consistent local
extremum for the computation within the region of physically relevant final
quadrupole moments.


\subsection{Parameter Space Volume}
\label{sec:param_vol}

The first step is to define the parameter space volume $\mathcal{V}$, from
which all designs $\mathcal{D}$ will be sampled. Our intent was that the
parameter space volume $\mathcal{V}$ should allow drawing identical designs for
the \UNEDFZERO, \UNEDFONE, and \UNEDFTWO\ protocols. Therefore, the volume was
chosen large enough for the statistical analysis to consider parameter values
beyond the 95\% confidence intervals around each of the previously determined
\UNEDFZERO, \UNEDFONE, and \UNEDFTWO\ optimization results but without being so
large that one sees many failed \HFBTHO\ computations, nonsensical results,
nonphysical results, or results with multiple possible solutions. Note that
some of the parameters of the energy functional (e.g., the nuclear
incompressibility $\knm$) are not very sensitive to ground-state properties and
are therefore ill-constrained: their most likely value determined by the
statistical calibration may be at the boundaries of the volume considered.

\Table{\label{tab:DesignBounds} 95\% confidence intervals (CIs) for the
{\UNEDFZERO}, {\UNEDFONE}, and {\UNEDFTWO} parameterizations; from
\cite{kortelainen2010,kortelainen2012,kortelainen2014}, respectively. When
available, the CI is rounded to the nearest integer except for $\rhosat$,
$\msca$, and $\enm$. The last column defines the volume studied in this work.
Note that the CIs for $C^{JJ}_t$ are given only for completeness, since these
coupling constants were set to 0 in this work.}
\br
Parameter              & \centre{1}{\UNEDFZERO} &   \centre{1}{\UNEDFONE} & \centre{1}{\UNEDFTWO} & Interval Studied Here \\
\mr
$\rhosat$              & $[\00.160,0.161]$ & $[\00.158,0.159]$ & $[\00.154,0.158]$   & $[\00.155,0.165]$\\
$\enm$                 & $[{-16.1},{-16.0}]$   &      \centre{1}{--}     &  \centre{1}{--}       & $[{-16.1},{-15.5}]$\\
$\knm$                 & \centre{1}{--}          &      \centre{1}{--}     & $[\m223,\m257]$       & $[\m200,\m245]$ \\
$\asym$                & $[\m\026,\m\036]$       &  $[\m\028,\m\030]$      & $[\m\029,\m\030]$     & $[\m\028,\m\032]$ \\
$\lsym$                & $[\0{-21},\m111]$       &  $[\m\022,\m\058]$      &  \centre{1}{--}       & $[\m\020,\m\060]$ \\
$1/\msca$              &    \centre{1}{--}       &  $[\0\00.9,\0\01.5]$    & $[\0\00.8,\0\01.2]$   & $[\0\00.8,\0\01.2]$ \\
$C_0^{\rho\Delta\rho}$ & $[\0{-58},\0{-52}]$     &  $[\0{-53},\0{-37}]$    & $[\0{-51},\0{-42}]$   & $[\0{-60},\0{-40}]$ \\
$C_1^{\rho\Delta\rho}$ & $[-149,\m\038]$         &  $[-218,\0{-73}]$       & $[-153,\0{-73}]$      & $[-160,\0{-50}]$ \\
$V_0^{\rm{n}}$         & $[-174,-166]$           &  $[-212,-160]$          & $[-223,-195]$         & $[-240,-150]$ \\
$V_0^{\rm{p}}$         & $[-205,-194]$           &  $[-225,-189]$          & $[-242,-219]$         & $[-265,-180]$ \\
$C_0^{\rho\nabla J}$   & $[\0{-85},\0{-74}]$     &  $[\0{-81},\0{-67}]$    & $[\0{-74},\0{-55}]$   & $[\0{-85},\0{-60}]$ \\
$C_1^{\rho\nabla J}$   & $[\0\0{-3},\m\094]$     &  $[\0{-68},\0\0{-4}]$   & $[\0{-64},\0{-13}]$   & $[\0{-80},\m\050] $\\
$C_0^{JJ}$             & \centre{1}{--}          & \centre{1}{--}          & $[\0{-82},\0{-24}]$   & $[-105,\m\010] $\\
$C_1^{JJ}$             & \centre{1}{--}          & \centre{1}{--}          & $[\0{-95},\0{-36}]$   & $[-120,\m\010] $\\
\br
\end{tabular}
\end{indented}
\end{table}
Table \ref{tab:DesignBounds} shows the intervals (last column) used to define
$\mathcal{V}$ and compares them with the 95\% confidence intervals (CIs) for
each of the three {\UNEDF} parameterizations. For the six coupling constants
related to the surface, spin orbit, and tensor terms of the Skyrme functional
and the two coupling constants of the pairing functional, the proposed interval
encompasses all three CIs (with the exception of $C_1^{\rho\nabla J}$ for
{\UNEDFZERO}). For the nuclear matter properties, physics constraints sometimes
impose tighter bounds than what the result of the statistical analysis may have
suggested.


\subsection{Self-Consistent Determination of Configurations for Deformed Nuclei}
\label{sec:design}

For simplicity, we assume that for a given nuclear configuration we can use the
same target quadrupole moment across all parameter space points $\xb$ in the
design $\mathcal{D}$. For example, in the case of spherical ground-state
nuclei, we adopt the accepted approach of setting $\qtarget(\xb) = 0$ b for all
$\xb$ in the design.  Note, however, that the constant value of $\qtarget$ can
be different for each nuclear configuration.

Given a volume $\mathcal{V}$ of the parameter space, we would like to determine
for each deformed configuration a $\qtarget$ value that is both motivated by
experimental results and representative of the final deformations for that
configuration across the entire volume. To this end, we devised a
self-consistent, iterative scheme with initial target values, $\qtarget^{(0)}$,
set by physical expectations. Specifically, for each deformed ground state,
we adopted an initial quadrupole moment that is consistent with an axial
quadrupole deformation $\beta_{2} = 0.3$; in the case of fission isomers, the
same configuration scheme is adopted but with the experimentally motivated
initial quadrupole moment set to the value consistent with $\beta_{2} = 0.6$.

Given a design $\mathcal{D}$ of points $\xb$ contained in the volume of
interest in the functional's parameter space, the procedures is as follows. For
each configuration in the study, the final values of the quadrupole moment
$\qfinal^{(i)}(\xb)$ at iteration $i$ are used to configure the observable
engine at iteration ${i+1}$. Specifically, for all points $\xb$ in the design
$\mathcal{D}$, we set $\qtarget^{(i+1)}$ to the median of the final quadrupole
moment values obtained for that state across all points $\xb$ in the design at
the previous iteration:
$\qtarget^{(i+1)} = \langle\qfinal^{(i)}(\xb)\rangle_{\mathcal{D}}$. A result
is included in the computation of the median only if it was convergent and if
postprocessing determined that the result is both sensible and physically
reasonable. This iterative procedure stops when no $\qtarget$ values change
appreciably for two successive iterations. For \UNEDF1, convergence was reached
in four such iterations with a maximum change in $\qtarget$ of approximately
0.05 b; see Table~\ref{tab:FailCounts}. We refer to this initial $\qtarget$
configuration as $\mathcal{C}_{0}$.

To search for other possible \HFBTHO\ solutions, we derived two other
$\qtarget$ configurations, $\mathcal{C}_{1}$ and $\mathcal{C}_{2}$, directly
from the results used to determine $\mathcal{C}_0$. For $\mathcal{C}_{1}$, the
target configuration for each state was set to the 12.5 percentile of the
$\qfinal$ values for $\mathcal{C}_{0}$; for $\mathcal{C}_{2}$, it was set at
the 87.5 percentile. These values are different enough to increase the
likelihood of finding other local extrema but close enough to avoid finding
solutions with final deformation too far from physically motivated expectations.


\subsection{Analysis of Acquired Data and Dependence on $\qtarget$}
\label{sec:q2_dependence}

\Table{\label{tab:FailCounts} Results of the self-consistent adjustment of
the target quadrupole moment $Q_{\rm t}$ for deformed nuclei. The first four
rows correspond to the iterations of the initial configuration
$\mathcal{C}_{0}$; the last two rows show the final results for two other
configurations; see text for details. Column 2 shows the maximum difference
between the target and final quadrupole moment across all nuclear
configurations; columns 3 and 5 are the percentage of nonconvergent and
nonphysical computations out of total 39,500 computations in the design,
respectively; column 4 is the number of nonsensical computations.}
\br
Config. &  max$(\Delta \qfinal)$ [b] &  Nonconv. &  Nonsens. & Nonphys. \\
\mr
$\mathcal{C}_0^{(0)}$ &  $33.279 $ &  1.87\% &  1 &    4.84\% \\
$\mathcal{C}_0^{(1)}$ &  $\01.767$ &  1.20\% &  1 &    0.90\% \\
$\mathcal{C}_0^{(2)}$ &  $\00.508$ &  1.12\% &  1 &    0.92\% \\
$\mathcal{C}_0^{(3)}$ &  $\00.051$ &  1.13\% &  1 &    0.93\% \\
$\mathcal{C}_1$       &  $\0{-}$   &  1.76\% &  0 &    0.90\% \\
$\mathcal{C}_2$       &  $\0{-}$   &  1.62\% &  1 &    0.96\% \\
\br
\end{tabular}
\end{indented}
\end{table}

The results of the self-consistent determination of the target quadrupole
moment configuration $\mathcal{C}_0$ are displayed in
Table~\ref{tab:FailCounts}.  We can clearly see that the first iteration, which
stepped the configuration away from one motivated purely by experimental
expectations, was successful in decreasing the number of failed computations
in the design. However, the table shows that even for the final datasets, the
number of nonconvergent computations and results with nonsensical or
nonphysical results is nonzero. We also note that for all three configurations
the number of parameter space points with at least one unacceptable result is
about half the total number of points in the design. In other words, for an
arbitrary $\xb$ point in the design, there is an approximately 50\% chance that
at least one of the 79 {\HFB} calculations needed to define the value of the
objective function (\ref{eq:chi2}) has failed. These failures seem unavoidable
and introduce an effective ``noise'' in the objective function. We  show later
that this noise has insignificant impact on the optimization and calibration.

\begin{figure}[!htb]
\begin{center}
\includegraphics[width=0.4\textwidth]{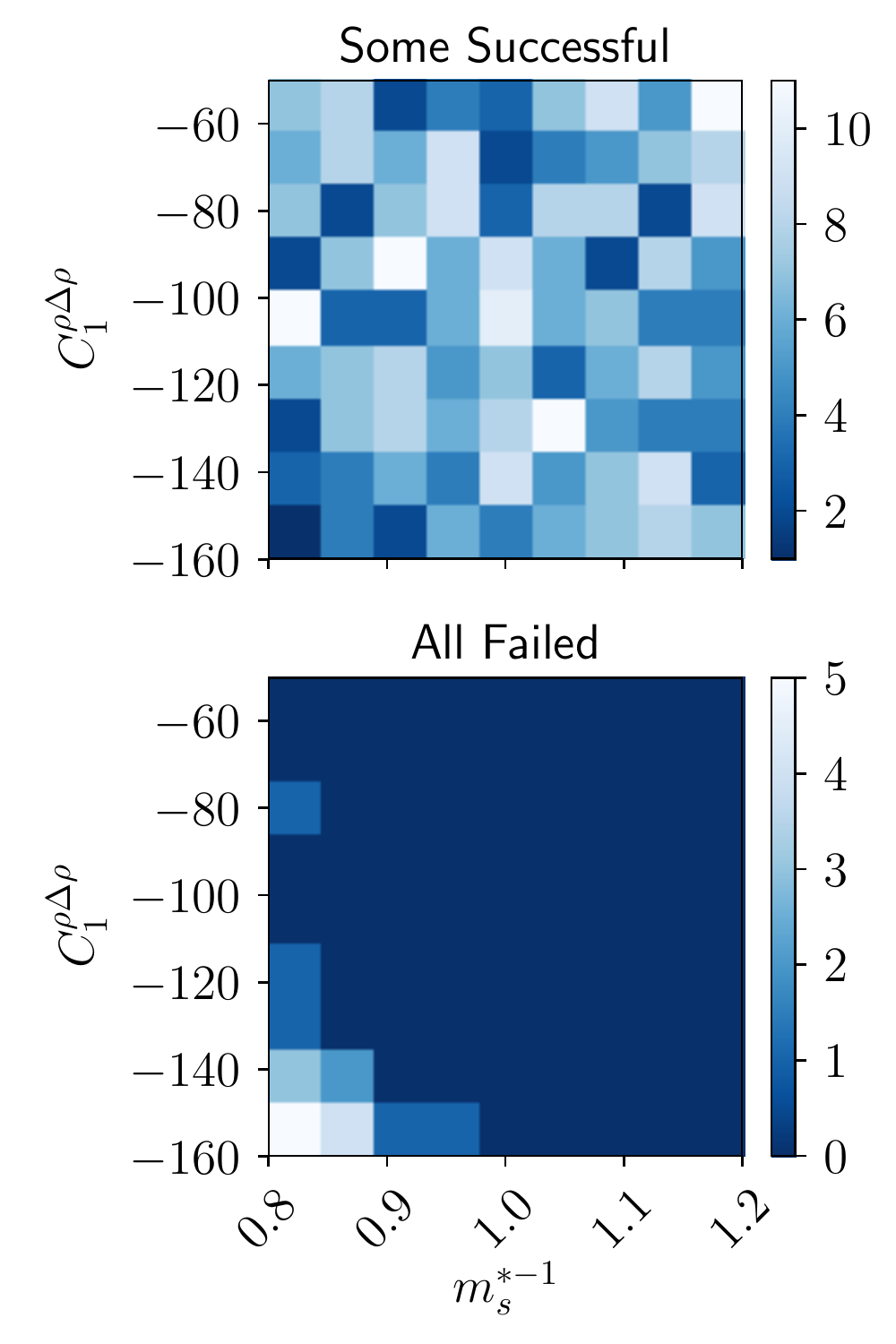}
\includegraphics[width=0.4\textwidth]{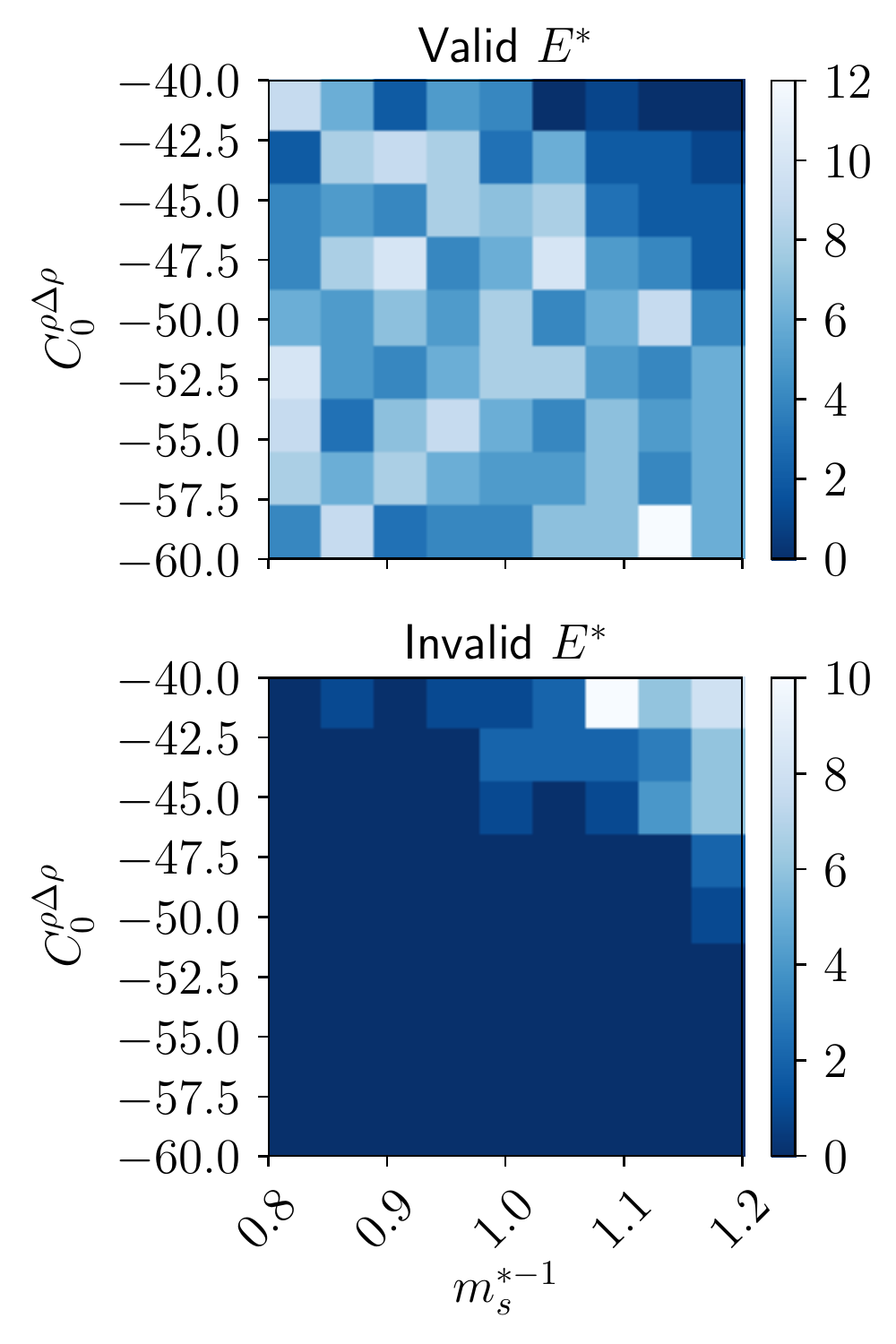}
\caption[]{Two-dimensional histograms showing the number of failures in a
two-dimensional projection of the $\nx$-dimensional space. Left column: failure
counts for ground-state binding energies, proton radii, and OES staggering in
the $(1/\msca, \CrDr{1})$ plane. The label ``All Failed'' indicates that at
least 3 different nuclear configurations had unacceptable results for all
configurations $\mathcal{C}_0, \mathcal{C}_1,$  and $\mathcal{C}_2$. Right
column: failure counts for fission isomer excitation energy in the
$(1/\msca, \CrDr{0})$ plane. The label ``Invalid $E^{*}$'' indicates that at
least 2 fission isomers were flagged.}
\label{fig:FailureRegions}
\end{center}
\end{figure}

The histograms in Figure~\ref{fig:FailureRegions} show that there is a region
of the parameter space volume in which invalid results are generated regardless
of the target quadrupole configuration used. There is also a second region in
which the binding energy of the fission isomer is significantly outside the
range of allowed values, for example, quite lower than the ground state.
Therefore, the failure statistics for the design $\mathcal{D}$ could be
significantly improved if such regions were excluded; we note, however, that
some failures would continue to be unavoidable even with such filtering.

Each combination of nuclear configuration and parameter space point in the
design was classified based on how many of the $\mathcal{C}_0, \mathcal{C}_1$,
and $\mathcal{C}_2$ computations failed and whether the valid results were
independent of the $\qtarget$ value used. For spherical ground-state
computations, we found that the difference in $\qfinal$ that arises from using
different $\qtarget$ is small overall, almost always less than 0.1 b.
Therefore, we deem them $\qtarget$-dependent if the maximal difference in
binding energy between valid computations exceeds 0.002 MeV. For deformed
ground state or fission isomer computations, the $\qfinal$ value for some of
the outliers can change by up to 10 b. The computation is thus classified as
$\qtarget$-dependent if the maximal difference in $\qfinal$ exceeds 0.5 b.

\begin{figure}[!htb]
\begin{center}
\includegraphics[width=1.0\textwidth]{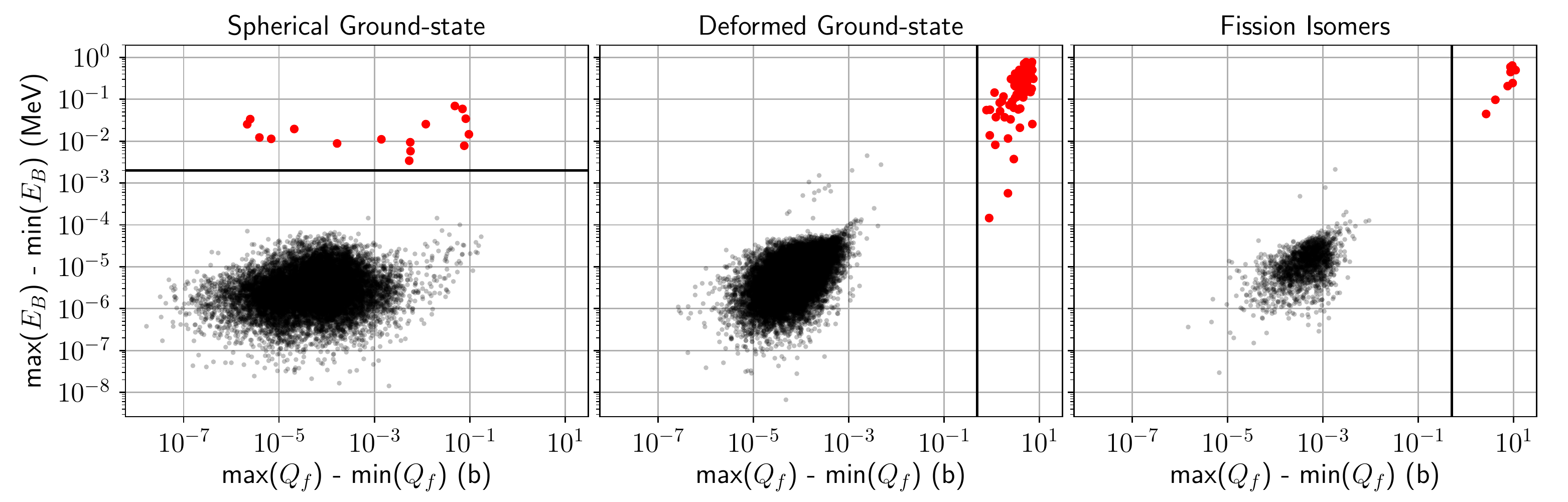}
\caption[]{Spread in energy plotted as a function of the spread in final
quadrupole moments across the three $\qtarget$ configurations discussed in
Section~\ref{sec:design}. }
\label{fig:QtThresholds}
\end{center}
\end{figure}

Figure~\ref{fig:QtThresholds} gives a visual representation of this analysis.
For the spherical and deformed ground-state energies, as well as for the
fission isomers, we recorded the spread in energy and final quadrupole moments
across the three $\qtarget$ configurations described in
Section~\ref{sec:design} when at least two of the $C_0, C_1, C_2$ results were
convergent, sensible, and physical. The spread values were computed only across
those results that were convergent, sensible, and physical. The figure shows
the spread in energy as a function of the spread in final quadrupole moment for
these three observables. Most of the computations are characterized by a very
small spread of a few dozen eV for the energy and on the order of millibarns
for the quadrupole moments. The outliers, marked in red and separated from the
rest by a line, are easily identified. Note that deviations can reach up to 1
MeV for the energy, which will introduce a larger contribution to the objective
function (\ref{eq:chi2}). The configurations with greater $\qtarget$ dependence
are often those  with many nonconvergent computations. The main conclusion of
this analysis is that the parameter space volume was indeed small enough that
the number of $\qtarget$-dependent results was relatively small.


\subsection{Effect on the Parameter Optimization}

To study the effect of $\qtarget$ configurations on optimization, we attempted
to reproduce the original {\UNEDFONE} optimization result reported
in~\cite{kortelainen2012}. Four optimizations, each using a different
$\qtarget$ configuration, were run for 350 $\xb$ evaluations; the {\POUNDERS}
optimization software from \cite{SWCHAP14} drove {\HFBTHO} via the observable
engine. These runs were set up such that each {\HFBTHO} computation used the
default {\HFBTHO} initial state corresponding to the given nuclear
configuration and $\qtarget$ value. As with the original {\UNEDFONE}
optimization, all optimization runs started from the {\UNEDFZERO} parameter
values and used the same bound constraints as those in~\cite{kortelainen2012}.

\begin{table}[!htb]
\lineup
\caption{\label{tab:OptimizationResults}History of {\UNEDFONE} optimizations. The
columns correspond to the parameter optimization constraints, the original
{\UNEDFONE} parameterization reported in Table II of~\cite{kortelainen2012},
the standard deviations also reported in Table II of~\cite{kortelainen2012},
and the parameterizations found by using the  $\qtarget$ configurations defined
here. The final four rows are the number of nonconvergent, nonsensical, and
nonphysical computations out of the total 27,650 computations in each
optimization, as well as the number of parameter space points with at least one
nuclear configuration whose computation was nonconvergent or yielded
nonsensical or nonphysical results. The parameter values in bold indicate that
the associated parameter was actively constrained by the bounds.}
\footnotesize\rm
\begin{tabular}{@{}*{8}{l}}
\br
{} & \centre{1}{Bounds} & \centre{1}{\UNEDFONE} & \centre{1}{$\sigma$} & \centre{1}{$\mathcal{C}_0$} & \centre{1}{$\mathcal{C}_1$} & \centre{1}{$\mathcal{C}_2$} & \centre{1}{$\mathcal{C}_0^*$} \\
\mr
$\rhosat$      & $[0.15, 0.17]   $       &$\m\0\00.15871$    & $ \00.00042$ & $\m\0\00.15850$    & $\m\0\00.15879$    & $\m\0\00.15881$    & $\m\0\00.15876$ \\
$\enm$         & $[-16.2,-15.8] $   &$\0\mathbf{-15.8}$ &\centre{1}{--}& $\0\mathbf{-15.8}$ & $\0\mathbf{-15.8}$ & $\0\mathbf{-15.8}$ & $\0\mathbf{-15.8}$ \\
$\knm$         & $[220, 260]     $   &$\m\mathbf{220}$   &\centre{1}{--}& $\m222.416$        & $\m220.000$        & $\m220.156$        & $\m220.340$    \\
$\asym$        & $[\028,\036]       $       &$\m\028.987$       & $\00.604$    & $\m\029.010$       & $\m\029.041$       & $\m\029.048$       & $\m\029.047$  \\
$\lsym$        & $[\040,100]      $       &$\m\040.005$       & $13.136$     & $\m\040.599$       & $\m\040.000$       & $\m\040.000$       & $\m\040.042$  \\
$\msca$        & $[\00.9,1.5]     $       &$\m\0\00.992$      & $\00.123$    & $\m\0\00.976$      & $\m\0\00.981$      & $\m\0\00.996$      & $\m\0\00.978$ \\
$\CrDr{0}$     & (-$\infty$, $\infty$)   &$\0{-45.135}$      & $\05.361$    & $\0{-44.064}$      & $\0{-44.636}$      & $\0{-45.131}$      & $\0{-44.370}$  \\
$\CrDr{1}$     & (-$\infty$, $\infty$)   &$-145.382$         & $52.169$     & $-140.159$         & $-136.479$         & $-136.506$         & $-136.847$   \\
$\VZeroN$      & (-$\infty$, $\infty$)   &$-186.065$         & $18.516$     & $-183.378$         & $-184.055$         & $-186.303$         & $-183.688$   \\
$\VZeroP$      & (-$\infty$, $\infty$)   &$-206.580$         & $13.049$     & $-204.971$         & $-205.605$         & $-207.136$         & $-205.146$   \\
$\CrDJ{0}$     & (-$\infty$, $\infty$)   &$\0{-74.026}$      & $\05.048$    & $\0{-73.007}$        & $\0{-73.585}$        & $\0{-73.663}$        & $\0{-73.193}$  \\
$\CrDJ{1}$     & (-$\infty$, $\infty$)   &$\0{-35.658}$      & $23.147$     & $\0{-28.553}$        & $\0{-28.431}$        & $\0{-32.673}$        & $\0{-30.990}$  \\ \hline
$\chi^2$       &        \centre{1}{--}   &$\m\052.201$       &\centre{1}{--}& $\m\051.942$       & $\m\051.920$       & $\m\051.967$       & $\m\051.890$  \\ \hline
{\small Nonconv.} &     \centre{1}{--}   &   \centre{1}{--}  &\centre{1}{--}& $\m306$            & $\m405$            & $\m421$            & $\m\030$       \\
{\small Nonsens.} &     \centre{1}{--}   &   \centre{1}{--}  &\centre{1}{--}& $\m\0\00$          & $\m\0\00$          & $\m\0\00$          & $\m\0\00$ \\
{\small Nonphys.} &     \centre{1}{--}   &   \centre{1}{--}  &\centre{1}{--}& $\m\0\02$          & $\m\0\00$          & $\m\0\00$          & $\m\0\00$ \\
{\small Failures} &     \centre{1}{--}   &   \centre{1}{--}  &\centre{1}{--}& $\m290$            & $\m283$            & $\m313$            & $\m\028$ \\
\br
\end{tabular}
\end{table}

As seen in Table~\ref{tab:OptimizationResults}, the parameterizations found
from these four optimization runs are consistent with the original {\UNEDFONE}
optimization results.  However, while $\knm$ was actively constrained at its
lower bound in the original study, this is not the case (although sometimes
$\knm$ was barely constrained) for any of the new parameterizations. Future
studies could include new {\UNEDFONE} optimizations run with relaxed bound
constraints on parameters such as $\enm$ and $\knm$ (e.g., using the intervals
that define the parameter space volume used in this study; see
\Cref{tab:DesignBounds}).

{Since our optimizations run without the benefit of the offline postprocessing
flagging of results mentioned in Section~\ref{sec:hfbtho}, the trajectories of
the optimization can potentially be affected by nonconvergent or nonphysical
results. The failure statistics provided in \Cref{tab:OptimizationResults} were
collected only after the optimizations finished and indicate that the
optimizations were robust to  nonconvergent {\HFBTHO} computations. The
majority of these failures occurred for the $(92, 146)$ fission isomer and only
after the $\chi^2$ value had decreased to a value close to the best one
reported here. An investigation of these failed computations revealed that
{\HFBTHO} was finding two possible solutions: the first was near an inflection
point in the {\pes} rather than an extremum (refer to
Section~\ref{sec:casestudy}), and the second was at the neighboring local
{\pes} maximum. It appears that computations tending toward the {\pes} maximum
have a much harder time converging within {\HFBTHO}.

To see whether an alternative initial state setup scheme would yield
``cleaner'' results, we used the fourth configuration $\mathcal{C}_0^*$. The
only difference between $\mathcal{C}_0^*$ and $\mathcal{C}_0$ is the scheme
used for setting the initial state for each {\HFBTHO} computation.
Specifically, for each nuclear configuration in the optimization protocol, the
first computation was run using the same default {\HFBTHO} initial state
mentioned above. For all subsequent computations, however, the initial state
was set using the solution for the same nuclear configuration from the previous
parameter space point. Although this optimization run encountered far fewer
nonconvergent computations, the parameterization obtained was not changed
significantly. The setup scheme of $\mathcal{C}_0$ should be preferable to that
of $\mathcal{C}_0^*$ since the solution of the latter has a potentially
stronger dependence on the initial starting point in parameter space as well as
the default initial state. That said, the fact that this optimization also
settles in the same region of the parameter space as the others bolsters the
case that {\POUNDERS} appears to find a consistent approximate local minimum
for this objective function.


\section{Statistical Emulation and Calibration }
\label{sec:calibration}

Bayesian inference has become common in the nuclear physics community for
quantification of uncertainties \cite{furnstahl2015,higdon2015,mcdonnell2015,
steiner2015,utama2016,utama2017,utama2018,niu2018,neufcourt2018,neufcourt2019,
niu2019}. The goal of Bayesian inference here is to estimate a distribution of
{\UNEDFONE}  parameters that probabilistically match experimental observations,
which are assumed to have been measured with error. To do so, we construct a
Bayesian model that has two components. The {\em likelihood}, $f(y|\xb)$, is
our distribution for the experimental observations $y$ given the unknowns
(i.e., the {\UNEDFONE} parameters $\xb$). The {\em prior}, $\pi(\xb)$, is a
marginal distribution for the unknown parameters that summarizes our knowledge
of these parameters before observing data. Following the rules of probability,
one can construct the posterior distribution of the unknown {\UNEDFONE}
parameters given the observations as
$$
p(\xb | y ) \propto \pi(\xb) f(y|\xb).
$$
One of the key ideas is that the likelihood $f(y|\xb)$ is based on forward
evaluations of the {\UNEDFONE} objective function in (\ref{eq:chi2}).

For complicated Bayesian models, the posterior distribution cannot be directly
integrated to obtain moments, and the normalization term cannot be directly
integrated to remove the  proportionality. Instead, Markov chain Monte Carlo
({\mcmc}) is typically used to draw samples from the posterior distribution.
{\mcmc} is a sequential sampling method that  requires forward evaluations only
of the unnormalized posterior distribution. For details on {\mcmc} methods, see
\cite{robert2005monte}.


\subsection{Statistical Emulation and Calibration of {\UNEDFONE}}
\label{sec:emu_cal}

Statistical quantities expressible as expectations calculated by using the
posterior samples will approximate the full posterior quantities with a
well-understood approximation error that shrinks as the number of {\mcmc}
samples increase. {\mcmc} methods are desirable approximations to the full
posterior for this reason; however, they require a large number of evaluations
of the likelihood for each evaluated parameter value. Because of high
correlation among some {\UNEDFONE} parameters, more than 1 million samples
would be required in order to get a reasonable estimation of posterior
quantities. That would require millions of evaluations of {\UNEDFONE} for all
experimental observables, which is an infeasible computational cost.

In order to overcome this computational bottleneck, the evaluation of
{\UNEDFONE} in the {\mcmc} process is replaced by a computationally inexpensive
emulator. In principle, any fast regression or function approximation technique
can be used as an emulator of {\UNEDFONE}. Gaussian processes  have been used
as emulators for many years across many fields, largely because of two main
advantages of {\gp} emulators. First, {\gp}s interpolate the observed values at
the locations where the computer model has been evaluated, properly reflecting
the information about the response of the computer model at that location.
Second, {\gp}s not only give flexible, accurate predictions but also provide an
estimate of uncertainty in predictions from the emulator at locations where the
model has not been evaluated. Thus, {\gp}s allow calibration and prediction to
properly account for limited evaluations of a computer model such as
{\UNEDFONE}.

A {\gp} is a stochastic process indexed by a $\nx$-dimensional input space
where any finite collection of random variables in the process are multivariate
normally distributed \cite{rasmussen2006}. The process defines a prior
distribution on functions of the input space,
\begin{equation}
f(\xb) \sim \mathcal{N}\big(\mu(\xb), k(\xb,\xb^\prime, \hyperp) \big) ,
\end{equation}
where $\mathcal{N}$ denotes a (multivariate) Gaussian distribution, $\mu(\xb)$
is the mean function of the process, $k(\xb,\xb^\prime, \hyperp)$ is a function
that gives the covariance between function outputs at two locations in input
space, and  $\hyperp$ is a vector of hyperparameters for the covariance
function (e.g., correlation length). A common covariance function is the
squared exponential function
$$
k(\xb,\xb^\prime, \hyperp) = \sigma \e^{-\frac{(\xb - \xb^\prime)^2}{\ell^2}} ,
$$
where $\hyperp = (\sigma,\ell)$ with $\sigma$ the marginal variance of the
process, which defines the scales over which functions in the function space
are expected to vary, and $\ell$ is the correlation length of the {\gp}. The
choice of covariance function and its hyperparameters dictates the properties
of the prior distribution on functions and the type of function space
supported~\cite{rasmussen2006}. The entry in the $i^{\rm th}$ row and
$j^{\rm th}$ column of the correlation matrix between outputs of $f(\xb)$ at a
finite set of locations in the input space is the correlation function
evaluated at $\xb_i$ and $\xb_j$, $k(\xb_i,\xb_j, \hyperp)$.

After defining the {\gp} prior, the {\gp} is updated by using Gaussian
conditioning on a set of evaluations of the computer model,
$\fb = [f(\xb_1),\ldots,f(\xb_{\ndes})]$. The {\gp} at a set of new locations,
$\xb^\ast$, conditioned on observation of $\fb$, that is,
$$
f(\xb^\ast) \mid \fb \sim \mathcal{N}\left( \tilde \mu(\xb^\ast), \tilde \Sigma(\xb^\ast,\xb^\ast)\right),
$$
has a posterior mean and covariance matrix of
\begin{eqnarray*}
\tilde \mu(\xb^\ast) & = & \mu(\xb^\ast) + \Sigma(\xb^\ast,\xb)\Sigma(\xb,\xb)^{-1} (\fb - \mu(\xb) ), \\
\tilde \Sigma(\xb^\ast,\xb^\ast) & =  &\Sigma(\xb^\ast,\xb^\ast) -  \Sigma(\xb^\ast,\xb)\Sigma(\xb,\xb)^{-1}
\Sigma(\xb,\xb^\ast),
\end{eqnarray*}
where $\Sigma(\xb,\xb)$ is the $\ndes \times \ndes$ covariance matrix with
entry $i,j$ equal to $k(\xb_i,\xb_j, \hyperp)$. The $\hyperp$-dependence in
$\Sigma$ has been suppressed for clarity. $\Sigma(\xb^\ast,\xb)$ represents the
$n_{\rm new} \times \ndes$ cross-covariance terms between outputs at the
$n_{\rm new}$ locations to be predicted by $f(\xb^\ast)$ and observed locations
of the function $f(\xb)$. The posterior mean, $\tilde \mu(\xb^\ast)$, is
recognized as an accurate emulator of many complex physical models
\cite{kennedy2001, kennedy2000predicting,higdon2004combining, higdon2008}. The
posterior variance represents uncertainty in the output of the function at a
location where it has not yet been evaluated. Inclusion of this uncertainty,
rather than using a point-estimate prediction, allows for uncertainty in a
function at unevaluated points to be reflected in the uncertainty associated
with estimating $\xb$.

For the emulation of {\UNEDFONE} in this work, the function (or computer model)
is the objective function (\ref{eq:chi2}). This function was evaluated at
$\ndes=500$ locations in the 12-dimensional volume $\mathcal{V}$ listed in
Table~\ref{tab:DesignBounds}. These locations form the design $\mathcal{D}$ and
were determined by a maximin, space-filling Latin hypercube design using the
\texttt{R} package \texttt{lhs} \cite{lhsPackage}. The choice of 500
evaluations was determined by balancing estimates of emulation error using
cross-validation with computational cost of larger sample size for building the
emulator.

Typically, during the calibration and emulation process, the {\gp}
hyperparameters are included in the sampling in order to fully account for
their uncertainty in the calibration process. Their inclusion requires that the
{\gp} covariance matrix across observations be rebuilt and inverted many times,
which can add substantial cost to the sampling process---the memory requirement
to store the covariance matrix scales as $\mathcal{O}(\ndes^2)$ and the
computational cost to invert scales as $\mathcal{O}(\ndes^3)$---and have minor
effect on the posterior distribution for the calibration parameters of
interest~\cite{liu2009modularization}. Instead, we follow the modularization
approach of Liu et al.~\cite{liu2009modularization} and fix the hyperparameters
to a point estimate---specifically to the maximum likelihood estimate using the
\texttt{scikit-learn} {\gp} implementation~\cite{scikit-learn}. Full sampling
via {\mcmc} of all parameters in the Bayesian model was carried out for 25,000
samples and showed posterior estimates consistent with the modularization
approach. Despite two weeks of computational time to collect those samples,
however, the Markov chains for the full Bayesian model were poorly converged
and thus not used in this work.

For emulation of (\ref{eq:chi2}), the input space for the {\gp} can be defined
both on the parameters of the model $\xb$ and on the physical inputs $\nub$
such as proton number $Z$ and neutron number $N$. In other words, we may write
the objective function as $f(\xb,\nub)$. Because {\HFBTHO} is run at all values
of $\nub$ for each $\xb$, however, the number of total observations would be
$\ndes \times n_d$, which can be computationally infeasible. Instead, we can
treat the vector of values, $[f( \xb,\nu_1), \ldots, f(\xb,\nu_{n_d})]$ as a
multivariate output of {\HFBTHO} rather than $n_d$ scalar outputs. The
evaluations of {\HFBTHO} can then be stacked into the $\ndes \times n_d$ matrix
$\mathbf M$.

To do so, we follow the approach presented in \cite{higdon2008,higdon2015},
using principal component analysis (PCA) to define a number of empirical basis
functions capturing the variation in the {\HFBTHO} output across $\nub$. Using
$n_b \ll n_d$ PCA bases, we can reconstruct the vector of {\HFBTHO} outputs
across $\nub$ as
$$
[m(\nub_1, \xb_i), \ldots, m(\nub_{n_d}, \xb_i)]
= \gras{\mu} + \mathbf S \sum\limits_{j=1}^{n_b} \mathbf k_j w_j(\xb_i)
= \gras{\mu} + \mathbf S \mathbf K\mathbf w,
$$
where $\mathbf K$ is a matrix made up of the PCA basis vectors $\mathbf{k}_j$
and $\mathbf w$ is the vector of PCA weights $w_j(\xb_i)$ as a function of the
parameters $\xb$. $\boldsymbol \mu$ is the $n_d$-length vector of the mean of
the columns of $\mathbf M$. $\mathbf S$ is an $n_d \times n_d$ diagonal matrix
of the empirical standard deviations of the same. For this study, twelve PCA
basis functions are used to capture 99.97\% of the variability of the output
across $\nub$. The weights, $w_j(\xb_i)$, can each be modeled with a {\gp}.

The resulting likelihood, given a parameter vector $\hat \xb$, is
$$
y \mid \hat \xb \sim \mathcal{N}\left(\mathbf S \mathbf K \tilde \mu(\hat \xb) + \gras{\mu}, \mathbf S \mathbf K \tilde \Sigma(\hat \xb,\hat \xb) \mathbf K^T \mathbf S  + \mathbf{\Lambda} \mathbf{\Sigma_\epsilon} \right),
$$
where $\mathbf{\Sigma_\epsilon}$ is the diagonal matrix capturing the given
measurement error and PCA truncation error. Experimental measurement error has
been provided with the observed data; however, a scaling matrix
$\mathbf{\Lambda}$ is introduced to allow the data to inform the precision with
which the {\UNEDFONE} model can capture variation in the experimental
observations. Following \cite{higdon2015}, we estimate a multiplicative scaling
factor $\lambda_j$ on the given standard deviation $\sigma_i$ of the
measurement error for each of the six data types: spherical nuclei binding
energy, deformed nuclei binding energy, proton {\rms} radii, proton pairing
gap, neutron pairing gap, and excitation energy of fission isomers. The
$i^{\rm th}$ diagonal entry of $\mathbf{\Lambda}$ corresponds to the
$\lambda_j$ for the data type of $y_i$.

Given this specification of the likelihood, the prior distributions for both
the {\UNEDFONE} parameters $\xb$ and scaling factors $\lambda_j$ are needed in
order to complete the statistical model. For $\xb$, upper and lower bounds on
physically plausible values in Table~\ref{tab:DesignBounds}  define the support
of the prior distributions. Within these bounds, the prior distribution was
determined to be uniform to express the assumption that all plausible values
were equivalently likely a priori. Each scale parameter $\lambda_j$ was
assigned a Gamma prior distribution, $\lambda_j \sim \Gamma(\alpha,\beta)$ with
$\alpha = \beta = 5$, reflecting the prior expectation that the multiplicative
scale should be near $\frac{\alpha}{\beta}$ = 1, but with a prior standard
deviation of $\frac{\sqrt{5}}{5}$, allowing reasonable variation near 1. As
stated before, the {\gp} hyperparameters were estimated from the simulation
results alone and then fixed.

To fit the Bayesian model, we obtain samples from the posterior distribution
using the ``No-U-turn'' variant of Hamiltonian Monte Carlo implemented in the
Stan probabilistic programming language~\cite{carpenter2017stan}. Four chains
of 1,500 samples each were obtained, and convergence of the chains to the
target posterior distribution was confirmed by using the $\hat R$
diagnostic~\cite{carpenter2017stan, gelman1992inference}.

\begin{figure}[!htb]
\includegraphics[width=\textwidth]{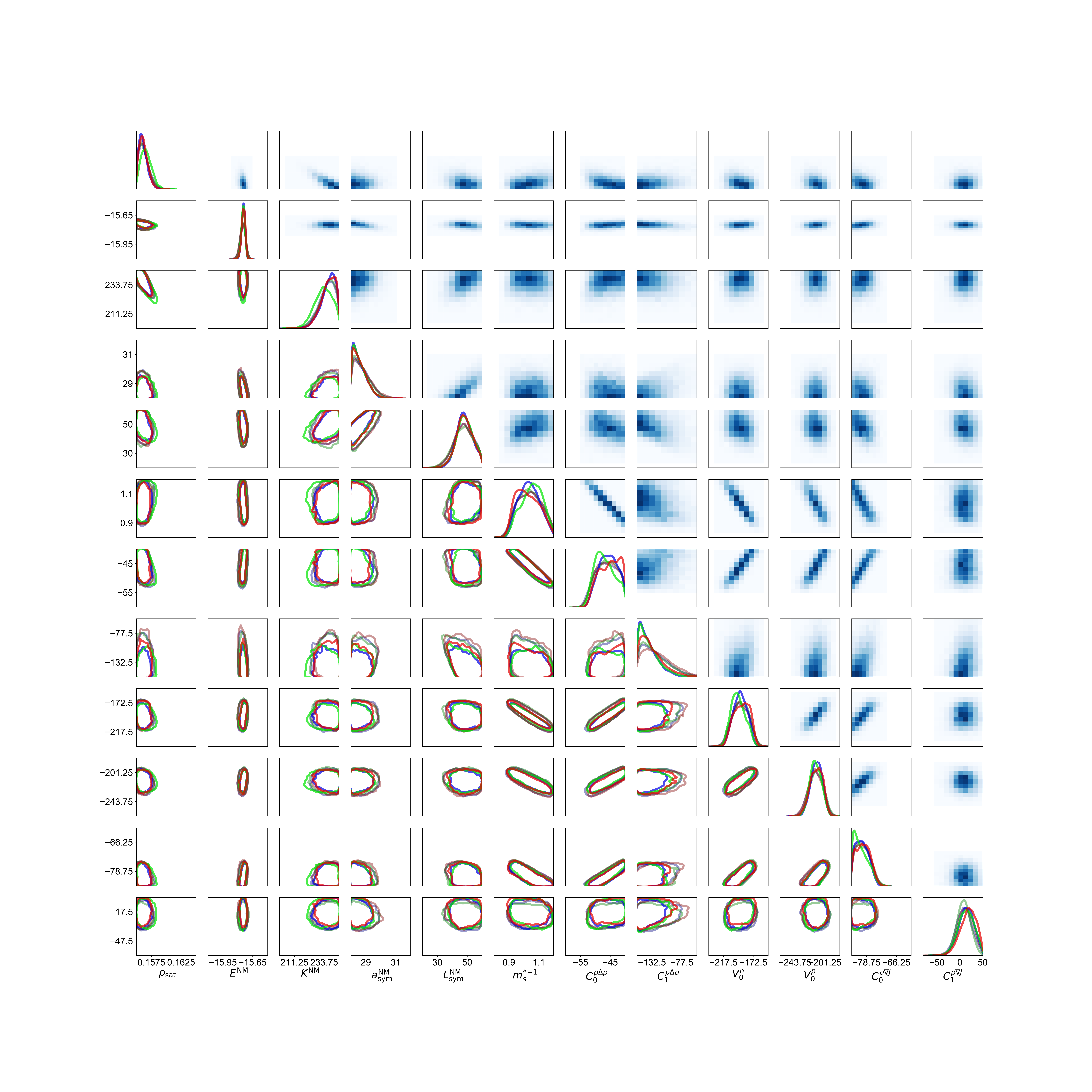}
\caption{Bivariate and univariate posterior summaries for the calibration
parameters. The diagonal shows the estimated 1D marginal posterior densities
for the three configurations $\mathcal{C}_0, \mathcal{C}_1,$  and
$\mathcal{C}_2$ with the muted color fit to the full {\UNEDFONE} data set and
the bright color fit removing runs with unacceptable observations (according to
the definition from Section~\ref{sec:q2_dependence}). The lower triangle
compares 2D joint 90\% credible regions. The upper triangle shows 2D histograms
for the 2D joint marginal distribution for $\mathcal{C}_0$ to illustrate the
structure of the highest posterior probability region.}
\label{fig:cal_compare}
\end{figure}


\subsection{Impact of Changing $\qtarget$ Target on Calibration of {\UNEDFONE}}

The statistical model described in Sec. \ref{sec:emu_cal} was fit with both the
full set of 500 evaluations of {\UNEDFONE} and a truncated set removing runs
with at least one unacceptable value as described in Sec.
\ref{sec:q2_dependence}. This resulted in 279, 238, and 244 out of 500 runs
included in the truncated set for $\mathcal{C}_0, \mathcal{C}_1,$  and
$\mathcal{C}_2$, respectively.

Comparison of the posterior distribution for the parameters $\xb$ shows little
evident sensitivity to the differing $\qtarget$ values for either the full or
truncated case. The diagonal and lower triangle of panels in
Figure~\ref{fig:cal_compare} show posterior summaries from the three $\qtarget$
configurations $\mathcal{C}_0, \mathcal{C}_1,$  and $\mathcal{C}_2$ for the
full and truncated models, with the color indicating the $\qtarget$
configuration. For a fixed color, the bright curve indicates the truncated data
and the muted curve indicates the full set. The diagonal panels show density
estimates of the 1D marginal distributions for each of the 12 parameters, while
the lower triangle of panels shows 90\% credible regions for the 2D joint
marginal distributions. For both the 1D and 2D marginals, the plots show
extremely high agreement across configurations, indicating little effect of
changing the $\qtarget$. This can also be quantified by using the {\mcmc}
$\hat R$ diagnostic~\cite{gelman1992inference}: values close to 1 for $\hat R$
indicate that the results are consistent with being one set of samples
originating from the same target distribution~\cite{gelman1992inference}. The
maximum $\hat R$ across all model parameters comparing chains from the three
full and truncated data cases were 1.0086 and 1.062 respectively, well below
the recommended threshold deviation from 1 of 1.1000~\cite{carpenter2017stan}.
The full and truncated results give consistent calibration in
Figure~\ref{fig:cal_compare}, with the truncated data leading to slightly more
concentrated posterior density for $\asym$, $\lsym$', and $\CrDr{1}$.

The upper triangle of panels in Figure~\ref{fig:cal_compare} shows 2D
histograms of the joint marginal distributions across parameters $\xb$ for
$\mathcal{C}_0$. Several parameters show strong pairwise correlations in the
posterior distributions: $\msca$, $\CrDr{0}$, $\VZeroN$, $\VZeroP$, and
$\CrDJ{0}$. The strong correlation is indicative of the parameter values being
only weakly identifiable from current data. Targeted measurements to
disentangle these correlations could substantially decrease the uncertainty of
all five parameters. Two other parameters, $\rhosat$ and $\knm$, show similar
high correlation. Because the posterior distribution is concentrated against
the boundary of the a priori feasible region, there is some evidence that
combinations of $\rhosat$ and $\knm$ may be consistent with the data that were
ruled out when choosing parameter ranges. This is also true of $\CrDr{0}$,
although $\CrDr{0}$ still has large uncertainty relative to the scale of the
prior range when compared with other parameters.

\Table{\label{tab:scaled_sigmas} Posterior mean and 95\% credible interval for
each of the 6 {\it a priori} theoretical errors $\sigma_{i}$ in (\ref{eq:chi2}).
The Bayesian model shows strong evidence that the data are consistent with a
standard deviation much smaller than that expected {\it a priori} for the
deformed nuclei binding energies, and with a slightly larger than expected
standard deviation for the fission isomer excitation energy.}
\br
& \multicolumn{2}{c}{Posterior estimates for $\sigma_i$} & \\
\mr
Data type & Mean & 95\% credible interval & Default\\
\mr
$E_{\rm sph}$ [MeV]    & 1.9500 & [1.7300, 2.6300] & 2.00 \\
$E_{\rm def}$ [MeV]    & 0.2270 & [0.2060, 0.2930] & 2.00 \\
$\Delta_{n}$ [MeV]     & 0.0457 & [0.0337, 0.0857] & 0.05 \\
$\Delta_{p}$ [MeV]     & 0.0703 & [0.0570, 0.1120] & 0.05 \\ 
$r_{p}$ [fm]           & 0.0177 & [0.0159, 0.0235] & 0.02 \\
$E^{*}_{\rm FI}$ [MeV] & 0.8500 & [0.7050, 1.3290] & 0.50 \\
\br
\end{tabular}
\end{indented}
\end{table}

In addition to the calibration of $\xb$, the scale $\sigma_{i}$ of the {\it a
priori} theoretical error was also informed by the data; see (\ref{eq:chi2}).
Table~\ref{tab:scaled_sigmas} shows the posterior mean and 95\% credible region
for each of the six data types for Bayesian model with the full data. The
posterior distribution for the standard deviation of binding energy for
deformed nuclei was much smaller than assumed, indicating that the data were
less variable and more informative for $\xb$ than thought {\it a priori}.
Conversely, the standard deviation for fission isomer excitation energy was
slightly larger, indicating more variable observed quantities than previously
expected. The estimated scale with the truncated data was consistent with
Table~\ref{tab:scaled_sigmas} and was omitted for brevity.


\section{Conclusion}

In this paper, we presented a comprehensive procedure to optimize and calibrate
nuclear energy density functionals when deformed nuclei are included in the
data. We paid special attention to the initialization of the self-consistent
calculations, which can have unwelcome impact on the characteristics of the
{\HFB} solution. While our case study was based on the {\UNEDFONE} Skyrme
functional, our results could easily be applied to the calibration of other
types of energy functionals. Our analysis leads to the following conclusions:
(i) Embedding diagnostic tools in the optimization/calibration process is
especially important to avoid regions of the parameter space that lead to
nonphysical solutions and to minimize the amount of ``noise'' in the computed
quantities such as the objective function (\ref{eq:chi2}). (ii) Owing to the
nonlinearity of the {\HFB} equations, some calculations will always fail one
way or the other during the optimization; but if the initial parameter space
volume has been well set up, our diagnostic tools show that these failures will
not dramatically impact the final parameterizations. (iii) Bayesian calibration
and direct optimization give  similar, robust (e.g., to the initializations
considered and code/dependency/compiler changes over the past several years)
results. (iv) By treating the a priori errors of each data type as
hyperparameters, Bayesian calibration can provide narrower estimates of these
errors. In the case of the {\UNEDFONE} functional, the estimate for the
standard error for deformed nuclei turned out to be 10 times smaller than
expected.

Our  estimate of theoretical error lend additional credence to the commonly
accepted view that deformed nuclei can be well described at the {\HFB}
approximation. The fact that the standard error for spherical nuclei is about
10 times larger than that for deformed nuclei also suggests that
beyond-mean-field corrections are needed to improve the overall quality of the
fit \cite{bender2006}. As  noted in \cite{kortelainen2014}, some of the
parameters of the Skyrme functional cannot be properly constrained by the
existing data and/or the limitations of the {\HFB} approximation. Although
changes in the initialization configurations of the calibration did not change
the values of the coupling constants ``much,'' the effect of these changes on
physics predictions needs to be tested on a case-by-case basis.


\section*{Acknowledgments}
This work was supported by the U.S.\ Department of Energy, Office of Science,
Offices of Advanced Scientific Computing Research and Nuclear Physics SciDAC
programs under Contract numbers DE-AC02-06CH11357 (Argonne) and
89233218CNA000001 (Los Alamos), and by the NUCLEI SciDAC project. It was partly
performed under the auspices of the US Department of Energy by the Lawrence
Livermore National Laboratory under Contract DE-AC52-07NA27344. We gratefully
acknowledge the computing resources provided by the Laboratory Computing
Resource Center at Argonne National Laboratory and by the Lawrence Livermore
National Laboratory Institutional Computing Grand Challenge program.


\section*{Bibliography}

\bibliographystyle{unsrt}
\bibliography{article}

\end{document}